# Electric control of the edge magnetization in zigzag stanene nanoribbon

Jingshan Qi[1*], Kaige Hu[2*], Xiao Li[3*]

[1]*School of Physics and Electronic Engineering, Jiangsu Normal University, Xuzhou 221116, People's Republic of China*

[2]*School of Physics and Optoelectronic Engineering, Guangdong University of Technology, Guangzhou510006, People's Republic of China*

[3]*Department of Physics, University of Texas at Austin, Austin, TX 78712, USA*

Abstract

There has been tremendous interest in manipulating electron and hole-spin states in low-dimensional structures for electronic and spintronic applications. We study the edge magnetic coupling and anisotropy in zigzag stanene nanoribbons, by first-principles calculations. Taking into account considerable spin-orbit coupling and ferromagnetism at each edge, zigzag stanene nanoribbon is insulating and its band gap depends on the inter-edge magnetic coupling and the magnetization direction. Especially for nanoribbon edges with out-of-plane antiferromagnetic coupling, two non-degenerate valleys of edge states emerge and the spin degeneracy is tunable by a transverse electric field, which give full play to spin and valley degrees of freedom. More importantly, both the magnetic order and anisotropy can be selectively controlled by electron and hole doping, demonstrating a readily accessible gate-induced modulation of magnetism. These intriguing features offer a practical avenue for designing energy-efficient devices based on multiple degrees of freedom of electron and magneto-electric couplings.

[*]Authors to whom correspondence should be addressed. Electronic addresses: qijingshan@jsnu.edu.cn, hukaige@gdut.edu.cn and lixiao150@utexas.edu

Since the first experimental realization of graphene,[1] research in two dimensional (2D) materials has experienced an explosive increase in recent years., Monolayer silicene,[2-4] germanene[5-7] and stanene[8,9] have also been recently synthesized. These group IV monolayers exhibit a variety of intriguing properties, such as band topology, Dirac states and valley physics. Spontaneous magnetism[10-12], together with topological edge states based on bulk-boundary correspondence[13-17], makes the study of their edges become an attractive field with potential application in energy-efficient quantum transport and nonvolatile memory. Room-temperature ferromagnetic order have been confirmed experimentally at the zigzag edge of graphene nanoribbons[10] and the magnetic edge anisotropy of two dimensional honeycomb systems has also been predicted by a single orbital Kane-Mele-Hubbard model, demonstrating an in-plane magnetization that opens a gap in the conducting edge channels.[18] Compared with graphene, the buckled geometry and considerable spin-orbit coupling in silicene, germanene and stanene may lead to distinct properties of edge states. On the other hand, the structure of 2D monolayer offers convenience to a gate modulation. The carrier doping can be easily introduced into 2D monolayer by a back-gate and may play an important role in determining the edge magnetism, which need to be explored.

In this letter, we investigate the edge magnetic coupling and anisotropy in zigzag edges of stanene monolayer by first-principles calculations. Given the spontaneous ferromagnetic order in each edge, both ferromagnetic (FM) and antiferromagnetic (AFM) couplings between two edges of the nanoribbon are studies, within-plane and out-of-plane magnetization rotation. We find that all zigzag stanene nanoribbons considered are all insulators when the spin-orbit coupling is included, and the band gaps depend on the magnetic order and magnetization direction. Distinct valley and spin degeneracies also appear with these magnetization variations. On the other hand, the magnetic coupling and anisotropy can be controlled by carrier doping. While the antiferromagnetic inter-edge coupling with an in-plane orientation is most stable in the absence of additional carrier doping, both electron and hole doping favor out-of-plane magnetization. Moreover, the hole doping also change magnetic coupling from AFM to FM. The gate-controllable edge magnetization can therefore realize a variable band gap and rich spin-valley physics, which pave an accessible way for novel electronic devices based on multiple electronic degrees of freedom and magneto-electric couplings.

The first-principles density functional theory[19, 20] calculations, as implemented in the VASP code[21, 22], are performed to relax the atomic structure and investigate electronic properties of zigzag stanene nanoribbons. The Perdew-Burke-Ernzerhof exchange-correlation functional[23] and the projector-augmented wave potentials[24] are used. An energy cutoff of 400 eV for the plane-wave basis and a 1×41×1 Monkhorst-Pack $k$-mesh[25] are adopted. A vacuum layer of more than 15Å thick is inserted to minimize the interaction between the nanoribbon and its periodic images. Atomic coordinates are optimized with a convergence threshold of 0.01 eV/Å on the interatomic forces. The magnetization is constrained along a certain direction by adding a penalty contribution to the total energy.

Fig. 1 shows atomic structures of a zigzag stanene nanoribbon, with three orthogonal reference axes $a$, $b$ and $c$. The $a$- and $b$-axes lie within the plane of the nanoribbon, respectively along the transverse and longitudinal directions of the nanoribbon, while the $c$-axis is along the out-of-plane direction. The zigzag nanoribbon considered here consists of 8 zigzag atomic chains, that is, sixteen tin atoms across the transverse direction in a unit cell. Each tin atom at outmost edges is saturated by a hydrogenatom, which restores the three-fold coordination of the tin atoms.[26] The zigzag nanoribbon with H-teminations is about 3.3 nm wide. Moreover, there are two tin atomic layers along $c$ axis with a 0.87 Å height difference, demonstrating a buckled structure, agreeing with previous results.[8, 9]

We then compute the edge magnetic anisotropy as the magnetization direction is rotated within $ab$-, $ac$-, and $bc$-planes. Given that the magnetic moment is localized at the edges of the nanoribbon and there is a ferromagnetic order in each edge, both FM and AFM couplings between two edges are taken into account in the calculations. Fig. 2(a) shows the total energy variation where the lowest energy is set to zero. AFM and FM inter-edge couplings have a similar change trend as rotating the magnetization, besides an energy shift of about 1 meV. For a certain magnetization direction, AFM coupling always has a lower energy than FM configuration, which is consistent with previous calculations[11] and recent experiments of graphene nanoribbon.[10] For both FM and AFM couplings, it is seen that the in-plane magnetization is remarkably more stable than the out-of-plane magnetization, demonstrating considerable magnetic edge anisotropy with an order of meV. In contrast, the energy change is smaller than 0.1 meV for in-plane magnetization rotation, with the magnetization along the *b* being the most stable configuration. Compared with the stanene nanoribbon, the graphene nanoribbon has a calculated anisotropic energy of less than 0.001 meV. The larger magnetic edge anisotropy in the stanene nanoribbon originates mainly from the stronger spin-orbit coupling of heavy element Tin. Besides, it is found that the magnetic coupling between two edges decreases gradually as the width increases by considering wider nanoribbons in our calculations, while the considerable anisotropy is well kept.

As changing the exchange coupling (AFM or FM) and rotating the magnetization direction, a few interesting feathers of electronic structures become compelling. Firstly, all magnetization states have band gaps and the magnitudes of these gaps depend on both the exchange coupling and the magnetization direction, as shown in shown in Fig. 2(b). The band gaps of AFM configurations are all more than 150 meV and increased with out-of-plane tilt of the magnetization. Compared with the AFM coupling, FM configuration always has a smaller band gaps for each magnetization direction. The band gap is about 70meV for the in-plane FM configurations, while the FM along the *c* axis has only a band gap of about 10meV. It is noted that the spin-orbit coupling is crucial for opening a band gap of the out-of-plane FM configuration. Only when the spin-orbit coupling is turned on in first-principles calculations, a band gap opens as shown in Fig. 3. This is very different from graphene nanoribbons, which is always metallic for out-of-plane magnetization.

On the other hand, depending on the magnetization, the spin and valley degree can been selectively expressed in the edge states. For the out-of-plane FM configuration, there is a large spin splitting of about 0.3 eV in the edge modes, as shown in Fig. 3(a).

Besides, the states at $K_+$ and $K_-$ in the momentum space are degenerate in energy, demonstrating a valley degeneracy. In contrast, the sizable spin-orbit coupling induces a valley splitting in the out-of-plane AFM configuration, where the energy of the highest occupied state at $K_+$ is 50 meV higher than that at $K_-$, while the spin degeneracy is kept, as shown in Fig. 4(b). Therefore, the distinct spin and valley degeneracies may provide an opportunity to selectively exploit the spin and valley degrees of freedom.

Taking one step from the above valley splitting, the spin degeneracy can further be broken by a transverse electric field in the nanoribbon with out-of-plane AFM configuration. Since the upward and downward magnetic moments are respectively located at left and right edges of the nanoribbon in Fig. 1 by the charge density analysis, spin-degenerate edge states are respectively localized at the two edges and polarized along the directions of corresponding magnetic moment. The spatial separation of the spin-degenerate states allow for a electric potential difference between two edges under a transverse electric field and associated spin-degeneracy breaking, as shown in Fig. 4(c) and (d). The direction of electric field determines the sign of the spin splitting. When the electric field is positive, namely, from left to right, the highest occupied state at $K_+$ becomes non-degenerate and the spin-up one from the left edge. When the electric field is reversed, the highest occupied state at $K_+$ is spin-down from the right edge. Therefore, spatially separated carriers with a certain spin and valley indices can be readily accessible by the in-plane electric field, which may has potential use in novel spintronic and valleytronic devices.

To realize these intriguing electronic properties with different magnetizations, the edge magnetic coupling and anisotropy need to be tuned in a controllable way. Given that the carrier doping can be easily introduced into 2D monolayer by a back-gate, we further calculate the carrier doping dependence of the magnetization. Fig.5 shows the anisotropic energies of both FM and AFM configurations, with the increase of electron/hole doping concentration. For the electron doping, while AFM is still favored over FM, it is seen that the total energy of out-of-plane magnetization becomes lower compared with in-plane magnetization as the doping concentration increases. A doping concentration of 0.05 $e$ per unit cell ($3.3 \times 10^{12}$ /cm$^{-2}$) makes the out-of-plane AFM configuration the most stable magnetic ground state (see Fig. 5(b)). On the other hand, the hole doping not only enhances the stability of out-of-plane magnetization states but also changes inter-edge magnetic coupling. A −0.05 $e$/cell doping makes the FM coupling more stable than the AFM coupling (see Fig. 5(d)). When the hole concentration is up to −0.25 $e$/cell, the out-of-plane magnetization becomes favored (Fig. 5(e)). Therefore, in-plane AFM, out-of-plane AFM, in-plane FM and out-of-plane FM configurations can all be achieved by carrier doping. Since the low-energy band structure is confirmed to be well kept with carrier doping besides the shift of the Fermi level, the intriguing electronic structures proposed above are selectively adopted by applying a back-gate.

Besides, we also consider edges terminated by halogen atoms, including fluorine, chlorine, bromine and iodine atoms. The magnetic edge anisotropy for these terminations is consistent with the hydrogen-terminated case, indicating that the magnetic edge anisotropy mainly origins from the spin-orbit coupling of the heavy element tin, regardless of the edge passivation.

In conclusion, we study edge magnetic coupling and anisotropy in zigzag edges of the stanene nanoribbon by the first-principles calculations. Owing to sizable spin-orbit coupling, zigzag stanene nanoribbon, demonstrates strong magnetic edge anisotropy. Depending on the types of inter-edge magnetic coupling (AFM and FM) and the magnetization directions, the band gaps vary within a large range, and intrinsic degrees of freedom of electron, valley and spin, exhibit distinct degeneracy. Especially for out-of-plane AFM configuration, two non-degenerate valleys of edge states emerge and the spin degeneracy is tunable by a transverse electric field, which give full play to spin and valley degrees of freedom. Furthermore, the magnetic coupling and anisotropy can be controlled by gate-induced carrier doping. Both electron and hole doping make the out-of-plane become more stable than in-plane. Hole doping can further change the magnetic coupling from AFM to FM. These intriguing features lead to high tunabilties of edge magnetization and associated electronic structures with the help of the carrier doping, which has potential application in advanced electronics based on multiple degrees of freedom of electron and magneto-electric couplings.

## Acknowledgments

J.Q.acknowledges the financial support from the National Natural Science Foundation of China (Projects No. 11674132) and PAPD. K. Hu acknowledges the financial support from the National natural Science Foundation of China (Projects No. 11647108) and the 100 Talents Program for Young Scientists of Guangdong University of Technology (Projects No. 220413139).

**Figures:**

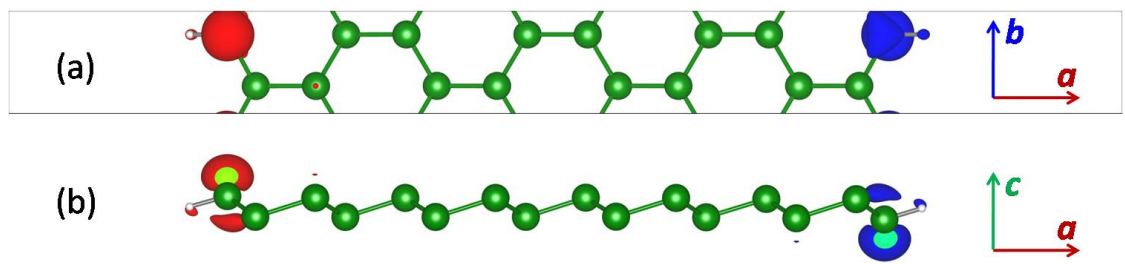

Fig.1. (a) Top view and (b) side view of the atomic structure of a stanene nanoribbon. The big green and small white balls denote tin and hydrogen atoms, respectively. The *a*- and *b*-axes are respectively along the transverse and longitudinal directions of the nanoribbon, while the *c*-axis is along the normal direction. The red and blue blobs represents the spin-up and spin-down charge density isosurfaces for AFM ground state, respectively.

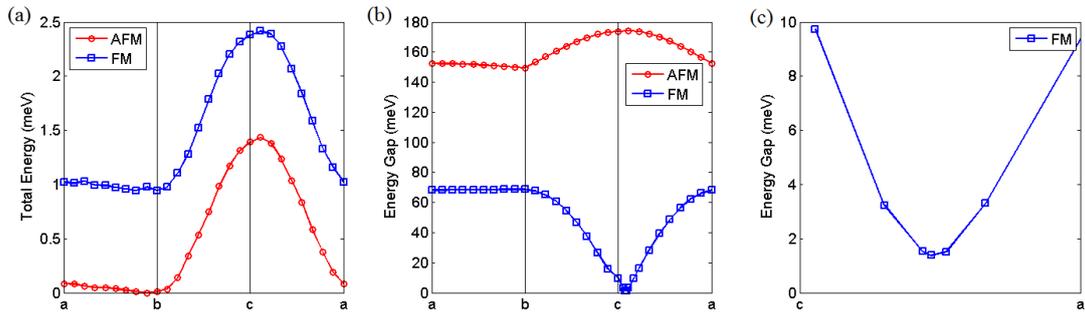

Fig.2. The evolution of (a) total energy and (b) band gap of the stanene nanoribbon with the magnetization rotation. Blue and red curves respectively correspond to inter-edge FM and AFM couplings. The total energy of the AFM configuration magnetized along *b* axis is minimum and set to zero. (c) zooms in the region near the *c*-axis with in the *ac*-plane, showing no band-gap closing.

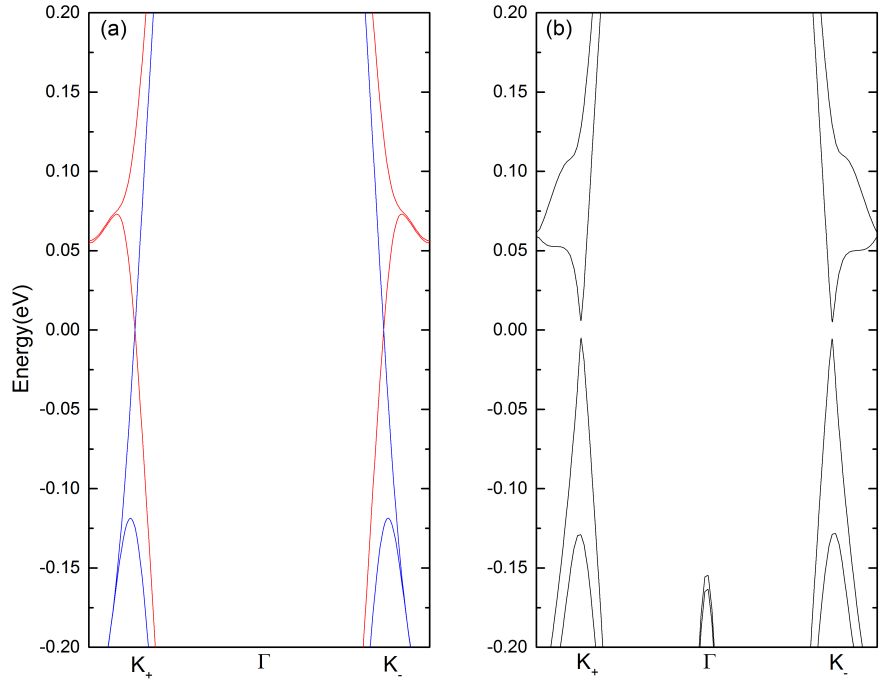

Fig.3. Band structures of out-of-plane FM configuration, (a) without and (b) with considering spin-orbit coupling. Red and blue lines in (a) respectively denote up- and down-spin bands.

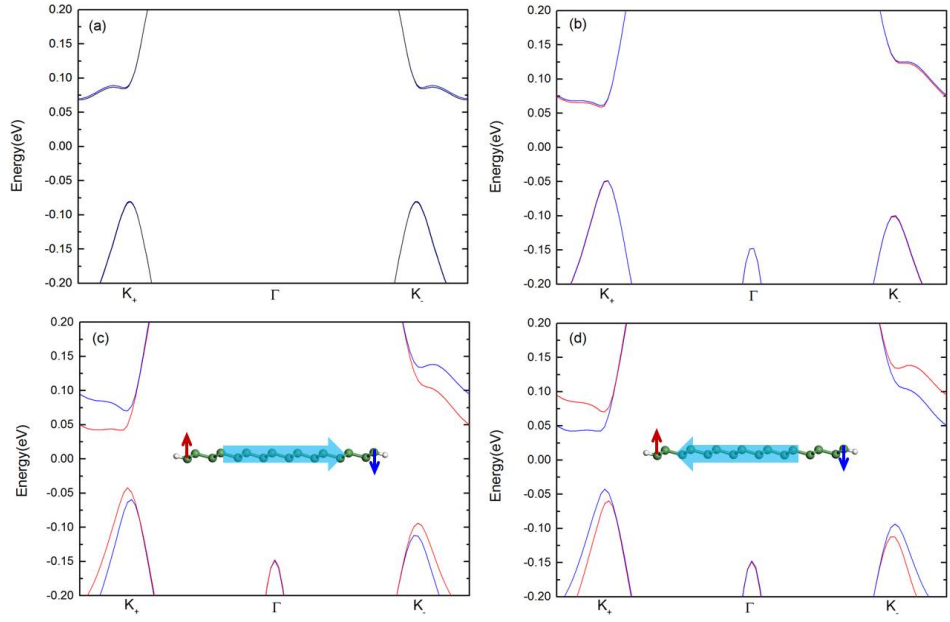

Fig.4. Valley and spin splitting for out-of-plane AFM configuration. Band structures (a) without and (b-d) with spin-orbit coupling. (c) and (d) are respectively applied a transverse electric field of 0.02 V/Å and -0.02 V/Å. Blue and red bands denote the down- and up-spin bands.

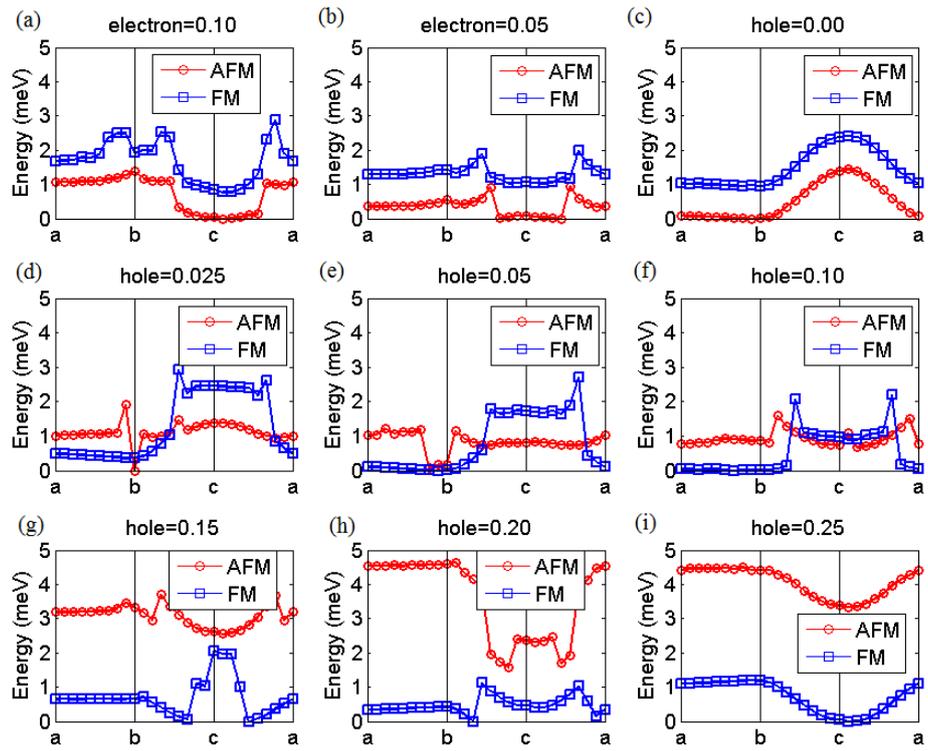

Fig.5. Total energy evolutions with the magnetization rotation when introducing electron/hole doping with different concentrations. The total energy of the most stable magnetization is set to zero.